\documentclass{aastex63}

\graphicspath{{./}{figures/}}

\submitjournal{ApJL}

\shorttitle{Lensing of Gravitational Waves}
\shortauthors{Rubin et al.}

\begin{document}

\newcommand{ \lensingdelay}{XXX~s\xspace}
\newcommand{\uhawaii}{\affiliation{Department of Physics and Astronomy, University of Hawai`i at M{\=a}noa, Honolulu, Hawai`i 96822, USA}}

\author[0000-0001-5402-4647]{D. Rubin}
\uhawaii
\affiliation{E.O. Lawrence Berkeley National Laboratory, 1 Cyclotron Rd., Berkeley, CA, 94720, USA}

\author[0000-0003-2274-0301]{Istv\'an Szapudi}
\affiliation{Institute for Astronomy, University of Hawai'i, 2680 Woodlawn Drive, Honolulu, HI 96822, USA}

\author[0000-0003-4631-1149]{B.~J.~Shappee}

\affiliation{Institute for Astronomy, University of Hawai'i, 2680 Woodlawn Drive, Honolulu, HI 96822, USA}

\author{Gagandeep S. Anand}
\affiliation{Institute for Astronomy, University of Hawai'i, 2680 Woodlawn Drive, Honolulu, HI 96822, USA}

\shortauthors{Rubin et al.} 
\title{Does Gravity Fall Down?\\Evidence for Gravitational Wave Deflection Along the Line of Sight to GW~170817}

\correspondingauthor{David Rubin}
\email{drubin@hawaii.edu}

\begin{abstract}

We present a novel test of general relativity (GR): measuring the geometric component of the time delay due to gravitational lensing. GR predicts that photons and gravitational waves follow the same geodesic paths and thus experience the same geometric time delay. We show that for typical systems, the time delays are tens of seconds, and thus can dominate over astrophysical delays in the timing of photon emission. For the case of GW~170817, we use a multi-plane lensing code to evaluate the time delay due to four massive halos along the line of sight. From literature mass and distance measurements of these halos, we establish at high confidence (significantly greater than $5 \sigma$) that the gravitational waves of GW~170817 underwent gravitational deflection to arrive within 1.7 seconds of the photons.

\end{abstract}


\section{Introduction} \label{sec:intro}

$1.734 \pm 0.054$ seconds after the LIGO/Virgo detection of the gravitational waves from GW~170817, Fermi and INTEGRAL observed the arrival of gamma rays \citep{multimessenger17, gravitationalwavesandgamma17}. This near-simultaneous arrival of gravitational waves and photons over 130 Myr of travel time provides a strict test for modified gravity theories \citep{baker17, langlois18}. \citet{boran18} estimate that the Shapiro delay \citep{shapiro64} was $\sim 400$ days (see also \citealt{gravitationalwavesandgamma17}), enabling further tests of general relativity (GR).

Recently, \citet{mukherjee19a} and \citet{mukherjee19b} have proposed measuring gravitational lensing of gravitational waves as a new probe of GR. They estimate that this lensing may be detectable in the future. In this work, we propose performing the same test, but using the geometric component of the lensing time delay as a test of GR. If gravitational waves and photons are both subject to the same geometric deflection, then they undergo the same lensing amplification. Likewise, if gravitational waves and photons undergo the same geometric deflection, then they should require the same travel time.\footnote{With a sample size of one (just GW~170817), it is extremely unlikely but perhaps possible that a conspiracy with different speeds for gravitational waves and photons could result in near-simultaneous arrival without gravitational waves undergoing the same geometric deflection as photons. This conspiracy would have to involve traveling through flat space (130~Myr), Shapiro delay ($\sim$~400 days) and the extra time due to geometric deflection of the photons ($\sim$~800 seconds, see Section~\ref{sec:GW}). This possibility will be completely eliminated if a second event, necessarily having a different relative combination of these travel times, also shows near-simultaneous photon and gravitational wave arrival.}

Investigating the possibility that gravitational waves undergo {\it larger} deflection than photons requires constraining the intrinsic time delay between gravitational-wave emission and photon emission. For example, gravitational waves could be emitted 100 seconds ahead of the photons, but delayed by 100 extra seconds due to larger deflection for near-simultaneous arrival. Considering this possibility is beyond the scope of this work.

Section~\ref{sec:timedelayformula} shows that, for typical nearby (tens of Mpc) gravitational wave sources, the geometric component of the time delay is of order tens of seconds. Section~\ref{sec:GW} shows our computation of an approximate time delay for GW~170817, obtaining a 68\% confidence interval of 400--2200 seconds. Thus, we show that the GW~170817 gravitational waves must have been geometrically deflected line-of-sight halos by an amount at least comparable to photons to have arrived at nearly the same time. We summarize and conclude in Section~\ref{sec:conclusion}.

\section{Single-Lens Time-Delay Estimate} \label{sec:timedelayformula}

\begin{figure*}
    \centering
    \includegraphics[width = 0.6 \textwidth]{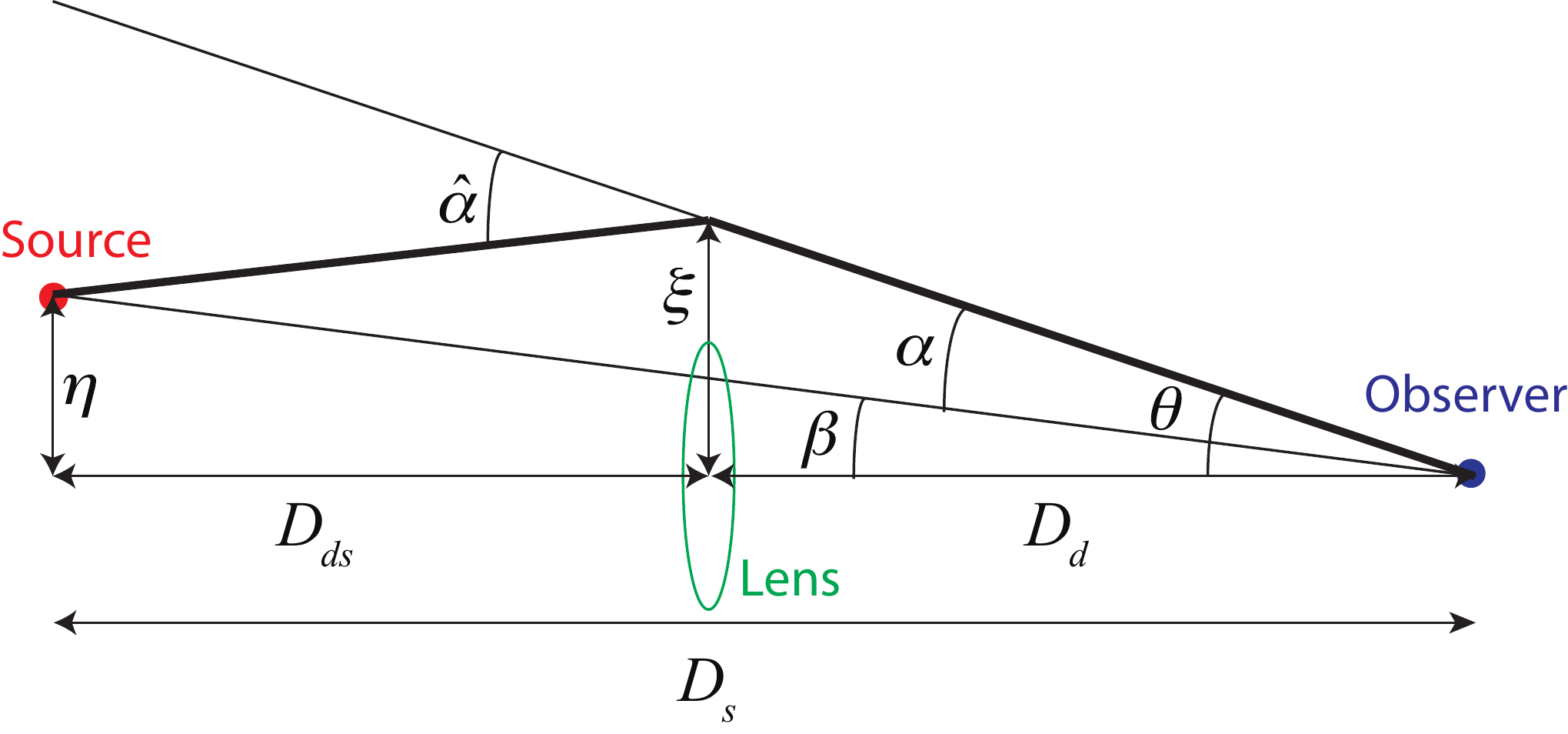}
    \caption{Geometry of a thin gravitational lens in the observer-source-lens plane using similar notation as, e.g., \citet{1992grle.book.....S}. The bold line shows the path the light takes from source to observer. $\hat{\alpha}$ is the deflection angle (Equation~\ref{eq:angle}), the impact parameter is $\xi$, the source is offset from the lens line of sight by a true (not observed) distance $\eta$, $D_d$ is the distance to the lens, $D_s$ is the distance to the source, and $D_{ds}$ is the distance from the lens to the source. For our derivation of a typical time delay in Section~\ref{sec:timedelayformula}, we assume all angles are small, so $\beta = \eta/D_s$, $\theta = \xi/D_d$, and $\alpha\; D_s = \hat{\alpha}\; D_{ds}$.
    \label{fig:illustration}}
\end{figure*}

We show an illustration of a single thin gravitational lens in Figure~\ref{fig:illustration}. For an axially symmetric thin lens (line-of-sight size much less than the line-of-sight distances), the lensing deflection is given by:
\begin{equation} \label{eq:angle}
    \hat{\alpha} = \frac{4 G M(\xi)}{c^2 \xi} = 1.91 \times 10^{-7}\ \mathrm{radians} \left[ \frac{M/(10^{12} M_{\odot})}{\xi / \mathrm{Mpc}} \right]  \;
\end{equation}

where $\xi$ is the impact parameter, and $M(\xi)$ is the enclosed mass at radius $\xi$ \citep[e.g.,][]{1992grle.book.....S}. For typical impact parameters of hundreds of kpc and galaxy-scale lenses, we can approximate $M(\xi)$ as the total mass $M$. For nearby sources like GW~170817 (redshift $\sim 0.01$), we can neglect the expansion of the universe (setting $z=0$). The extra path length due to the geometric deflection from the lens is
\begin{equation} \label{eq:pythagorean}
    \sqrt{D_{ds}^2 + (\xi - \eta)^2} + \sqrt{D_d^2 + \xi^2} - \sqrt{D_s^2 + \eta^2}\;.
\end{equation}

Assuming small angles so that we can substitute $\beta = \eta/D_s$, $\theta = \xi/D_d$, $\alpha\; D_s = \hat{\alpha}\; D_{ds}$, and also using $\alpha + \beta = \theta$, we expand Equation~\ref{eq:pythagorean} to lowest order in $\hat{\alpha}$ and $\theta$ to find the extra path length is

\begin{equation}
    \frac{1}{2} \frac{D_d D_{ds}}{D_s} \hat{\alpha}^2 \;.
\end{equation}

Thus the geometric component of the time delay is

\begin{equation} \label{eq:timedelay}
    \frac{1}{2 c} \frac{D_d D_{ds}}{D_s} \hat{\alpha}^2 = 18.9 \ \mathrm{seconds}\left[ \frac{D_d D_{ds}}{D_s} \frac{1}{10 \mathrm{Mpc}} \right] \left[ \frac{M/(10^{12} M_{\odot})}{\xi / \mathrm{Mpc}} \right]^2  \;.
\end{equation}

These time delays are of the order of the astrophysical time delay between gravitational waves and photons and will thus frequently be detectable, depending especially on $M$ and $\xi$ (both of which enter quadratically).

\section{Constraints from GW~170817} \label{sec:GW}

To estimate the geometric time delay, we select galaxies with possible lensing contributions along the line of sight to GW~170817 from the 2MASS Redshift Survey \citep{huchra12}. For each of the 43,533 2MASS galaxies included in the survey, we estimate the time delay up to a multiplicative constant from Equation~\ref{eq:timedelay}. We estimate the distance from the CMB-centric redshift, the impact parameter from the distance and angular separation, and the mass from the absolute $K$-band isophotal magnitude. \citet{coulter17} discovered the GW170817 optical emission, necessary to obtain the coordinates on the sky and the identity of the host galaxy (NGC~4993). \citet{cantiello18} provide a precise distance to NGC~4993: $40.7^{+2.5}_{-2.3}$~Mpc. Four galaxies are the most plausible for large time delays: NGC~5084, M104, M83, and NGC~5128 (Centaurus~A). The later two of these are the central galaxies in groups and all have dynamical halo mass measurements.\footnote{An alternative to measuring total halo mass is to estimate the stellar mass through the luminosity and then use a measured halo mass/stellar mass relation \citep[e.g.,][]{behroozi10} to estimate the halo mass. However, it is difficult to estimate masses to better than a factor of two with this approach, with roughly half coming from the stellar mass estimate \citep{conroy13}, and half from the uncertainties of and scatter around halo mass/stellar mass relations \citep{more09, behroozi10}.  This leads to at least a factor of four uncertainty in time delay, as time delay scales as (mass)$^2$. Furthermore,  when estimating through absolute magnitudes there is an additional dependence on distance of stellar mass~$\propto$~(distance)$^2$, as the distance is necessary to infer absolute magnitudes.} To be conservative, we assume all dynamical mass measurements are at fixed distance. As estimated dynamical mass scales with estimated distance, in addition to the mass uncertainties discussed below, we assume additional uncertainty on mass that covaries with the distance uncertainty.

With a dynamical mass as large as $10^{13} M_{\odot}$ \citep{carignan97}, NGC 5084 is one of the most massive disk galaxies known. For a more conservative mass measurement, we use the \citet{carignan97} value excluding two possible satellites that may not be bound: $(5.2\pm2.9)\times 10^{12} M_{\odot}$, which we convert to a log$_{10}(M/M_{\odot})$ measurement of $12.72 \pm 0.24$. We take a distance modulus of $31.12\pm 0.54$ (or 16.7 Mpc, \citealt{springob14}), giving an impact parameter of $0.84 \pm 0.24$~Mpc. We note that for this measured distance, there is a $\sim 0.02\%$ chance that NGC~5084 is actually behind NGC~4993 and thus NGC~5084 could not lens GW~170817.

M104 has a lower mass than NGC~5084, but it is more precisely measured. \citet{tempel06} find $2\times10^{12} M_{\odot}$, in good agreement with the \citet{jardel11} value for 50 kpc. \citet{jardel11} measure to larger scales and find a higher mass ($\sim 3\times10^{12} M_{\odot}$ with about 10\% uncertainty). To be conservative, we use $(2 \pm 0.2) \times10^{12} M_{\odot}$. We average two consistent tip of the red-giant branch distance measurements to M104 (\citealt{mcquinn16} and the Extragalactic Distance Database; \citealt{jacobs09}) to arrive at distance modulus of $29.88 \pm 0.08$ (or 9.46 Mpc), giving an impact parameter of $2.27 \pm 0.08$ Mpc.

\citet{karachentsev07} compute tip of the red-giant branch distances and the halo masses for the NGC~5128 and M83 groups. They find a mean distance of $3.76 \pm 0.05$ Mpc for the NGC~5128 group (for an impact parameter of $1.31 \pm 0.02$ Mpc) and $4.79 \pm 0.1$ Mpc for the M83 group (impact parameter of $0.74 \pm 0.02$ Mpc). Their mass for the NGC~5128 group is (6.4--8.1)$\times 10^{12} M_{\odot}$, which we convert to a log$_{10}(M/M_{\odot})$ measurement of $12.86 \pm 0.05$. For the the M83 group, they find (0.8--0.9)$\times 10^{12} M_{\odot}$, which we convert to a log$_{10}(M/M_{\odot})$ measurement of $11.93 \pm 0.02$. Of course, modeling these groups as axially symmetric masses (for the purposes of Equation~\ref{eq:angle}) is incorrect in detail. However, we show below that the implied time delays are large enough that moderate changes to our assumptions will not affect our conclusions.

\subsection{Combined Result}

\begin{figure}[]
    \centering
    \includegraphics[height=0.9\textheight]{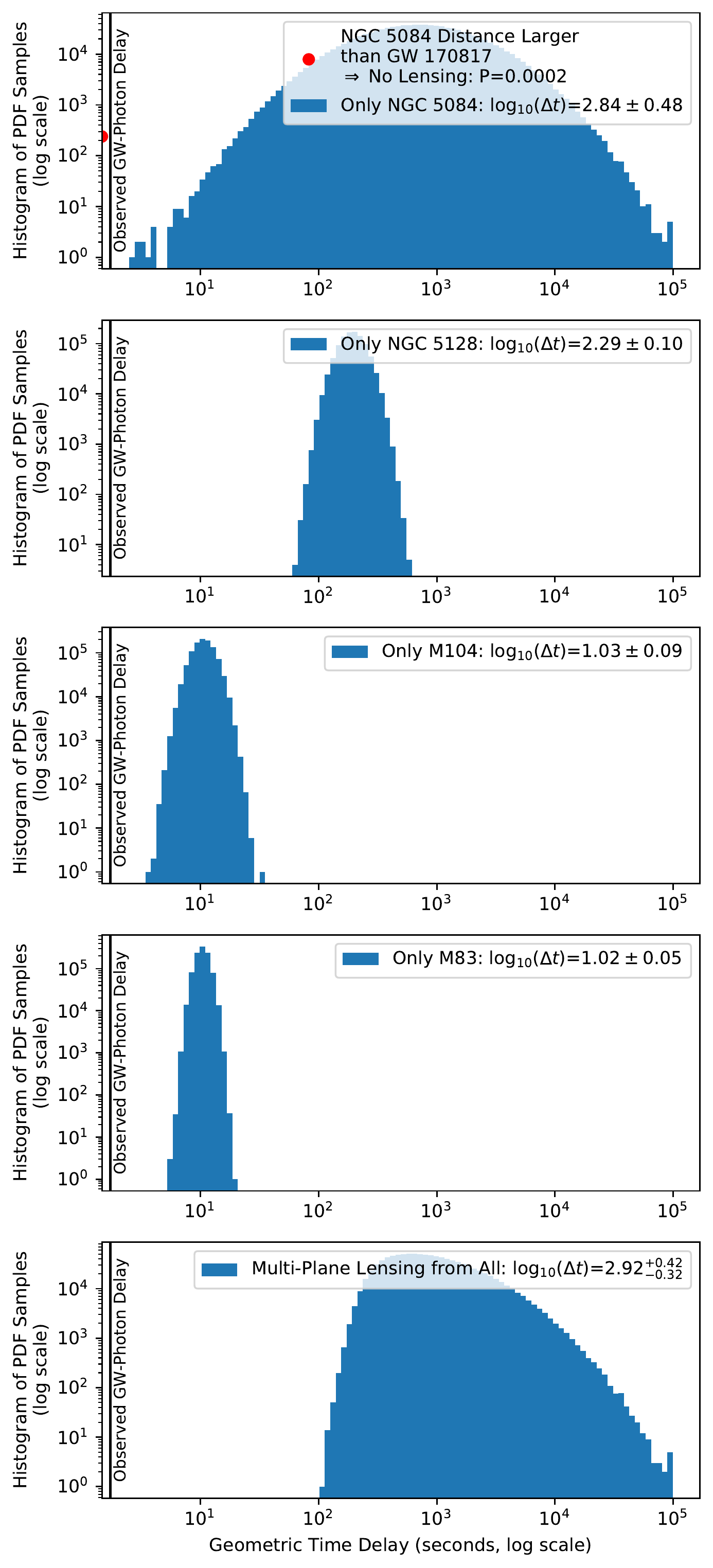}
    \caption{Constraints on the geometric component of the time delay along the line of sight to GW~170817. We draw 1,000,000 realizations, varying halo masses and distances, then compute the time delay for each. As the constraints are nearly log-normal, we make histograms of the time delays in log-spaced bins. The {\bf top four panels} show the time delays for each halo; the {\bf bottom panel} shows the multi-plane lensing calculation for all halos. To investigate the tails of the distributions, we show the y-axis logarithmically. We show the observed time delay between gravitational waves and photons with a vertical line. Each of the 1,000,000 computed time delays for the multi-plane combined result is well above the observed time delay, indicating a secure detection of the geometric component of the time delay for the gravitational waves.
    \label{fig:pdf}}
\end{figure}

To properly evaluate the total geometric delay due to each of these four halos along the line of sight, we wrote a simple multi-plane lensing code. This code starts a ray at the position (relative to us) of GW~170817 propagating towards the origin. As the ray reaches the position of closest approach to each halo, it is deflected according to Equation~\ref{eq:angle}. Due to the deflections during propagation, the ray will not pass through the origin. We thus iterate, adjusting the initial direction in the next iteration to account for the miss in the last iteration. As the deflections are $\sim 10^{-7}$ radians, this iteration converges rapidly. The difference between an undeflected line from GW~170817 and the path taken gives the time delay. As this relative difference scales as the deflection angles squared ($\sim 10^{-14}$) and the precision limit for double-precision (64 bit) floating-point is $\sim 10^{-16}$, we improve the accuracy of our results by using higher precision arithmetic through the Python decimal package.

We compute a the probability density function for lensing by numerically propagating all distance and mass uncertainties. We draw 1,000,000 realizations for all distances and masses and compute the geometric component of the time delay for each realization. The resulting probability density functions are shown in Figure~\ref{fig:pdf}. The top four panels show the time delay considering each halo in isolation; the bottom panel shows the multi-plane combination. Except for NGC~5084 (which does not have a precise enough distance measurement to securely place it along the line of sight to GW~170817), we see strong ($> 5 \sigma$) evidence of a time delay larger than the observed delay for each of the other halos. We see even stronger evidence for the combination (bottom panel of Figure~\ref{fig:pdf}) where the minimum time delay out of the realizations is 112 seconds and the 68\% confidence interval is 400--2200 seconds.

We can also interpret our result in terms of constraints on the deflection angle of gravitational waves ($\hat{\alpha}_{\mathrm{GW}}$), assuming that the gravitational waves left no later than the photons and traveled at the same speed. Starting from the lower bound of our 99.9999\% confidence interval (112 seconds), the gravitational waves were deflected by at least 110 seconds. Assuming Equation~\ref{eq:angle} applies to photons, we find
\begin{equation}
\hat{\alpha}_{\mathrm{GW}} > \frac{3.96 G M(\xi)}{c^2 \xi} \;.
\end{equation}

\section{Conclusion} \label{sec:conclusion}

This work proposes a novel test of GR: detecting the deflection of gravitational waves by a gravitational potential by evaluating the geometric component of the time delay due to the deflection. Any deflection more or less than predicted by GR will lead to a different time delay with respect to the photons. We show that typical galaxy halos give detectable time delays (of order tens of seconds) for gravitational-wave sources as close as tens of Mpc. We present an initial evaluation of the deflection the gravitational waves of GW~170817 underwent due to four halos along the line of sight. We see strong evidence that deflection did occur and GR passes our test.

\acknowledgments

This research has made use of the NASA/IPAC Extragalactic Database (NED),
which is operated by the Jet Propulsion Laboratory, California Institute of Technology, under contract with the National Aeronautics and Space Administration. IS acknowledges support from the NSF, award 1616974. The authors also thank Harald Ebeling and Brent Tully for feedback on this work. B.J.S. is supported by NSF grants AST-1908952, AST-1920392, and AST-1911074. We thank the anonymous referee for valuable feedback that substantially improved this work.

\software{
Matplotlib \citep{matplotlib}, 
Numpy \citep{numpy}, 
Python,
SciPy \citep{scipy}}

\end{document}